\documentclass[10pt]{article}

\usepackage{amsmath}
\usepackage{amssymb}  
\usepackage[T1]{fontenc}
\usepackage{pifont}
\usepackage{pstricks}
\usepackage{amsfonts}
\usepackage{graphicx}
\usepackage{amsmath}
\usepackage{authblk}
\usepackage{pstricks} 
\usepackage{pstricks-add}
\usepackage{caption}
\usepackage{placeins}
\usepackage{pstricks}
\usepackage{pdfpages}
\usepackage{framed}
\usepackage{cancel}
\usepackage{bm}
\usepackage{epstopdf}
\usepackage{tikz}
\def\rr2dot{\mathop{\bf r}\limits}
\def\x2dot{\mathop{x}\limits}
\def\y2dot{\mathop{y}\limits}

\def\bfy2dot{\mathop{\bf y}\limits}
\def\z2dot{\mathop{z}\limits}

\def\csi2dot{\mathop{\xi}\limits}
\def\et2dot{\mathop{\eta}\limits}
\def\bet2dot{\mathop{\beta}\limits}
\def\t2dot{\mathop{\theta}\limits}
\def\s2dot{\mathop{\sigma}\limits}
\def\d2dot{\mathop{\delta}\limits}
\def\q2dot{\mathop{q}\limits}
\def\l2dot{\mathop{\lambda}\limits}
\def\ps2dot{\mathop{{\cal E}}\limits}
\def\tet2dot{\mathop{\theta}\limits}
\def\bfx2dot{\mathop{\bf x}\limits}
\def\bfy2dot{\mathop{\bf y}\limits}
\def\bfq2dot{\mathop{\bf q}\limits}
\def\bfr2dot{\mathop{\bf r}\limits}
\def\bbfq2dot{\mathop{\bar {\bf q}}\limits}
\def\w2{\mathop{W}\limits}
\def\xgrande2dot{\mathop{\bf X}\limits}
\def\p02dot{\mathop{P}\limits}
\def\a2dot{\mathop{A}\limits}

\newtheorem{prop}{Proposition}
\newtheorem{cor}{Corollary}
\newtheorem{propr}{Property}
\newtheorem{rem}{Remark}

\title{Equations of motion for general nonholonomic systems from the d'Alembert principle via an algebraic method}
\author{F.~Talamucci}
\affil{{\it DIMAI, Dipartimento di Matematica e Informatica ``Ulisse Dini''},\\
{\it	Universit\`a degli Studi di Firenze, Italy}\\
{\it	e-mail: federico.talamucci@unifi.it}}
\date{}

\begin{document}
	\bibliographystyle{plain}
	
	\setcounter{equation}{0}

	\maketitle
	
	\vspace{.5truecm}
	
	
	\vspace{.5truecm}
	

	\vspace{.5truecm}
	

\begin{abstract}
	
\noindent	
The aim of this study is to present an alternative way to deduce the equations of motion of general (i.~e.~also nonlinear) nonholonomic constrained systems starting from the d'Alembert principle and proceeding by an algebraic procedure. 
The two classical approaches in nonholonomic mechanics --  ${\check {\rm C}}$etaev method and vakonomic method -- are
treated on equal terms, avoiding integrations or other steps outside algebraic operations.
In the second part of the work we compare our results with the standard forms of the equations of motion associated to the two method and we discuss the role of the transpositional relation and of the commutation rule within the question of equivalence and compatibility of the  ${\check {\rm C}}$etaev and vakonomic methods for general nonholonomic systems.
\end{abstract}

\section{Introduction}

\noindent
Our study concerns discrete mechanical systems subject to constraints involving the coordinates of the points and their velocities.
We refer to this situation as nonholonomic systems even if the denomination, more generally, encompasses all situations complementary to the case of purely geometric restrictions, i.~e.~only on the coordinates of the points and possibly the time. Our approach is theoretical, that is, we deal with the mathematical formulation of the model and not with the physical realization of the constraints.

\noindent
A fundamental point of reference for a comprehensive and historical review and for a systematic exposition of nonholonomic mechanics is \cite{neimark}. For an update in the following decades and for an important analytical treatment of nonholonomic systems we quote \cite{papa}. 
The classical treatment of nonholonomic systems regards in the majority of cases linear constraints, i.~e.~the constraint equations are linear with respect to the kinetic variables.
Among recent works that face the most general case of nonlinear kinematic constraints  we refer to \cite{benenti} for a comprehensive and general method for writing the dynamical equations through which various examples are analysed. In \cite{benentipendulum} an example of a nonlinear system is presented, together with a review of the theory. A second series of works studying nonholonomic nonlinear constrained systems can be found in \cite{zek1}, \cite{zek2}, \cite{zek3}. 

\noindent
Our attention turns to constraints of a general type: one of the main objectives of the work is to explore the possibility of applying the d'Alembert principle to nonlinear nonholonomic systems, using elementary algebraic techniques. By ``algebraic'' we essentially intend to avoid the use of integrals with respect to time, a step very often recurring in most formulations.
This possibility is known to exist for holonomic systems, constrained with only geometric restrictions.
The idea to extend the d'Alembert principle to general nonholonomic systems requires only the definition of the virtual displacements (i.~e.~possible displacements) for which the virtual work of the constraint forces vanishes (ideal displacements).

\noindent
In order to establish what types of displacements to consider, it is certain that we have to refer to the two main approaches prevailing in the theory of general constrained systems: we mean the displacements verifying the  ${\check {\rm C}}$etaev condition, historically traceable to \cite{cetaev} on one side, and displacements complying the ``vakonomic condition'', \cite{kozlov1}, \cite{kozlov2} \cite{kozlov3} (an interesting historical summary of the method is contained in \cite{lemos}) for which we find in literature only integral (and not algebraic) approaches (this is indeed a motivation for the work).

\noindent
The first part of the work (Section 2) is dedicated to recalling the D'Alembert principle in the most appropriate form and we develop simple linear algebra problems, coming from the declaration of two different classes of virtual displacements (class $(A)$ and class $(B)$).
The main result is stated in Proposition 1, where it is shown that the two systems of equations corresponding to the two different classes $(A)$ and $(B)$ are equivalent, if the displacements conditions verify a specific assumption (stated in (\ref{a=c})).
We opt to start from assigning physical coordinates to the model and to develop the theory by using the radius vector of each point; then we show (Paragraph 2.4) that by passing to any arbitrary set of Lagrangian coordinates the formal structure of the problem and the results are identical.

\noindent
In Section 3 the specific assignment (\ref{rc}) of the displacements conditions in terms of the constraint functions identifies the class $(A)$ with the method based on the ${\check {\rm C}}$etaev condition, class $(B)$ with the formulation of the vakonomic mechanics. Proposition 3 which transfers the results of Section 2 to the mentioned application states that the two methods are equivalent if both the displacements conditions are assumed, as far as the equations of motion are obtained through algebraic considerations.

\noindent
The final discussion in Section 4 compares the two set of nonholonomic equations (originating from $(A)$ and $(B)$) with those present in literature and  classified as ${\check {\rm C}}$etaev systems or vakonomic systems. In particular, the vakonomic algebraic method is checked against the procedure of deducing the equations from a variational principle (Hamilton--Suslov principle).
As a further issue, the possibility of making the conditions of the two kind  valid simultaneously is explored.
The question of equivalence of the two methods is indeed a current matter of debate and recently
various works are dedicated to the study of the compatibility (or inequivalence) of the two methods  (\cite{flanelusive} \cite{lemos}, \cite{lizamm}, \cite{liromp}, \cite{zampieri}).

\noindent
A significant role in the question is played by the so called transpositional relation, which establishes a link between the two conditions on the displacements (${\check {\rm C}}$etaev and vakonomic types), in terms of the lagrangian derivatives of the constraint functions and of a quantity which vanishes if the commutation property between displacement and velocity is assumed (in simple terms: whether the velocity of the displacement is equal to the displacement of the velocity).
Such a rule is another debated issue, starting from \cite{neimark} up to the significant and exhaustive
discussion in \cite{flan}. 
If on the one hand the validity of the commutation ule (tacitly or explicitly assumed in most of the vakonomic methods) entails the equivalence with the  ${\check {\rm C}}$etaev method only for holonomic systems (\cite{flanenigma}, \cite{lemos} \cite{lewis}), on the other hand the hypothesis of a non--zero coomutation rule offers new substance and significance to the vakonomic model (\cite{llibre}, \cite{pastore}).
The final observations of the work are developed precisely on the role of the transpositional relation assuming  the simultaneous validity of the displacements conditions. Our comments are based mainly on the general vakonomic method presented in \cite{llibre} and developed in (\ref{pastore}).

\section{Statement of the model}

\subsection{D'Alembert principle and virtual displacements}

\noindent
We consider a mechanical system consisting of $N$ material points  whose position is identified by the ${\Bbb R}^{3N}$ position vector ${\bf r}=({\bf r}_1, \dots, {\bf r}_N)$, and subject to the nonholonomic constraints (linear or nonlinear)

\begin{equation}
	\label{vinc}	
	\Psi_\nu ({\bf r}, {\dot {\bf r}}, t)=0, \qquad \nu=1, \dots, \kappa <3N
\end{equation}
where ${\dot {\bf r}}=({\dot {\bf r}}_1, \dots, {\dot {\bf r}}_N)$ is the velocity vector of the system.
The constraints are independent in the sense that
\begin{eqnarray}
	\label{psiindep} \textrm{the}\;\kappa\;\textrm{vectors} \;\dfrac{\partial \Psi_\nu}{\partial {\dot {\bf r}}}\in {\Bbb R}^{3N},\; \nu=1, \dots, \kappa &\textrm{are linearly independent}
\end{eqnarray}
or, equivalently, the rank of the $(\kappa\times 3N)$ jacobian matrix $\dfrac{\partial {\bm \Psi}}{\partial {\dot {\bf r}}}$, ${\bm \Psi}=(\Psi_1, \dots, \Psi_\kappa)$ achieves its maximum value $\kappa$.

\noindent
The Newton equations of motions for the systems can be written as

\begin{equation}
	\label{newton}
	{\dot {\bf Q}} = {\bf F}({\bf r}, {\dot {\bf r}},t)+{\bm R} 
\end{equation}
with ${\bf Q}=(m_1 {\dot {\bf r}}_1, \dots, m_N {\dot {\bf r}}_N)$ is the linear momentum of the system ($m_i$ is the mass), ${\bf F}= ({\bf F}_1, \dots, {\bf F}_N)$ lists the forces acting on the points and ${\bf R}=({\bf R}_1, \dots, {\bf R}_N)$ are the unknown forces due to restrictions (\ref{vinc}).
The equations become operational if the mechanical behavior of the constraints is specified: following the d'Alembert Principle of virtual works, the constraints forces reveal the property
\begin{equation}
	\label{ic}
	{\bm R} \cdot \delta {\bf r}=0
\end{equation} 
for the virtual displacements $\delta {\bf r}=(\delta {\bf r}_1, \dots, \delta {\bf r}_N)$ of the system (i.~e.~the displacements performed at a fixed time anfd consistent with the constraints).

\noindent
Equations (\ref{newton}) together with conditions (\ref{ic}) provide
\begin{equation}
	\label{dap}
	\left(	{\dot {\bf Q}} - {\bf F} \right)\cdot \delta {\bf r}=0
\end{equation}
from which the correct equations of motion will be deduced.
The question therefore shifts to identifying the appropriate $\delta {\bf r}$: the selection must be compatible with the constraints (\ref{vinc}) i.~e.~it must be expressed in terms of the functions $\Psi_\nu$; if on the one hand in the case of holonomic constraints (that is in absence of ${\dot {\bf r}}$ or in case of integrable constraints) the answer is clear and unambiguous, in the case of general constraints we can say that the question is open, especially in the nonlinear case.

\noindent
From a formal point of view for the moment, we indicate two general categories of displacements:

\begin{itemize}
	\item[$(A)$] the displacements verify 
	\begin{equation}
		\label{delta1}
	A({\bf r}, {\dot {\bf r}},t)\delta {\bf r}={\bf 0}
		\end{equation}
	where $A$ is a $\kappa \times 3N$--matrix with full rank $\kappa$. 
\item[$(B)$] the displacements fulfil the condition
\begin{equation}
	\label{delta2}
	B ({\bf r}, {\dot {\bf r}},t)\delta {\bf r}+C({\bf r}, {\dot {\bf r}},t)\delta {\dot {\bf r}}
	={\bf 0}
\end{equation}
where $B$, $C$ are $\kappa \times 3N$--matrices and $\delta {\dot {\bf r}}$ are the virtual variations of ${\dot {\bf r}}$ consistent with (\ref{vinc}).
\end{itemize}

\subsection{The mathematical problem}

For the moment we are dealing with an abstract situation: the model must be completed by linking the matrices $A$, $B$ anc $C$ to the constraint functions (\ref{vinc}). The problem is simply posed in these terms: once the position and the velocity of the system are fixed (by ${\bf r}$ and ${\dot {\bf r}}$), which displacements are compatible with the conditions (\ref{delta1}) or (\ref{delta2})?

\noindent
Case $(A)$ can  be expanded by a simple argument of linear algebra: for each fixed ${\bf r}$, ${\dot {\bf r}}$ and $t$ the solution of (\ref{delta1}) is the totality ${\cal W}$ of $\delta {\bf r}$ orthogonal to the row vectors ${\bf A}_\nu\in {\Bbb R}^{3N}$ of the matrix $A$, $\nu=1, \dots, \kappa$:
	$$
	\delta {\bf r}\in{\cal W}=\langle {\bf A}_1, \dots, {\bf A}_\kappa \rangle^\perp
	$$ 
Since the vectors are linearly independent, ${\cal W}$ is a vector space of dimension $n-\kappa$.
	Owing to (\ref{ic}), the constraint force ${\bf R}$ is orthogonal to all the vectors of ${\cal W}$, hence it must be ${\bf R}\in {\cal W}^\perp= \langle  {\bf A}_1, \dots, {\bf A}_\kappa \rangle$, namely 
	${\bf R}=\sum\limits_{\nu=1}^\kappa \lambda_\nu {\bf A}_\nu$ for some coefficients. We conclude that  equations (\ref{newton}) joined to (\ref{ic}), where the displacements have to verify (\ref{delta1}), are equivalent to
	\begin{equation}
		\label{eq1}
		{\dot {\bf Q}}={\bf F}+\sum\limits_{\nu=1}^\kappa \lambda_\nu {\bf A}_\nu\qquad case\;(A)
	\end{equation} 
where the multipliers $\lambda_\nu$ are unknown quantities. The essential assumption is the full rank of $A$. Equations (\ref{eq1}) are coupled with (\ref{vinc}) in order to form a system of $3N+\kappa$ equations in the unknowns ${\bf r}(t)\in {\Bbb R}^{3N}$ and $\lambda_1$, $\dots$, $\lambda_\kappa$. The essential final step (missing for the moment) is to establish the class (\ref{delta1}) according to the restrictions (\ref{vinc}), namely to set $A$ as a function of ${\bf \Psi}_\nu$ and their derivatives.

\noindent
Let us now examine case $(B)$: assuming that the rank of $B$ is full, we see that (\ref{dap}) and (\ref{delta2}) entail 
\begin{equation}
\label{eq2a}
{\dot {\bf Q}}-{\bf F}=\sum\limits_{\nu=1}^\kappa \mu_\nu {\bf B}_\nu + {\bf y}
\end{equation}
where the vector ${\bf y}$ is subject to the request
\begin{equation}
\label{wdr}
{\bf y}\cdot \delta {\bf r}=\sum\limits_{\nu=1}^\kappa \mu_\nu{\bf C}_\nu \cdot \delta {\dot {\bf r}}
\end{equation}
(${\bf B}_\nu$, ${\bf C}_\nu$ are the column of $B$, $C$ and $\mu_\nu$ are unknown mutliplying factors). We transform  the previous expression by considering that
$$
\mu_\nu {\bf C}_\nu \cdot \delta {\dot {\bf r}}=\dfrac{d}{dt}(\mu_\nu{\bf C}_\nu \cdot \delta {\bf r})-
\dfrac{d}{dt}(\mu_\nu {\bf C}_\nu)\cdot \delta {\bf r}+\mu_\nu {\bf C}_\nu \cdot \left(\delta {\dot {\bf r}}-\dfrac{d}{dt}\delta {\bf r}\right)
$$
The presence of the last term is due to the uncertainty of the rule (we will deal with this question afterward)
\begin{equation}
	\label{dtdelta}
	\dfrac{d}{dt}\delta {\bf r} =\delta {\dot {\bf r}}
\end{equation}
which would express the commutation of the operations $\frac{d}{dt}$ and $\delta$ (obviously ${\dot {\bf r}}=\frac{d}{dt}{\bf r}$). 

\noindent
Let us assume that it is possible to write (the hypothesis will be discussed later)
\begin{equation}
	\label{w}
	\delta {\dot {\bf r}}-\dfrac{d}{dt}\delta {\bf r}=W ({\bf r}, {\dot {\bf r}}, \bfr2dot^{..},t)\delta {\bf r}
\end{equation}
where $W$ is a square matrix of order $3N$. In that case, (\ref{wdr}) reduces to 
$$
{\bf y}\cdot \delta {\bf r}=
\sum\limits_{\nu=1}^\kappa 
\left(
-\dfrac{d}{dt}(\mu_\nu {\bf C}_\nu)+\mu_\nu W^T {\bf C}_\nu\right) \cdot \delta {\bf r}+
\dfrac{d}{dt}\left( \sum\limits_{\nu=1}^\kappa \mu_\nu {\bf C}_\nu \cdot \delta {\bf r} \right)
$$
so that (\ref{eq2a}) can be written as

\begin{eqnarray}
	\label{eq2}
	{\dot {\bf Q}}-{\bf F}&=&\sum\limits_{\nu=1}^\kappa \left( \mu_\nu {\bf B}_\nu 
	-\dfrac{d}{dt}(\mu_\nu {\bf C}_\nu)+\mu_\nu W^T {\bf C}_\nu \right)+{\bf y}_1\\
	&=&
\nonumber
\sum\limits_{\nu=1}^\kappa \mu_\nu \left( {\bf B}_\nu - \dfrac{d}{dt}({\bf C}_\nu)
	+ W^T {\bf C}_\nu\right)
	-\sum\limits_{\nu=1}^\kappa {\dot \mu}_\nu {\bf C}_\nu+{\bf y}_1\qquad case\;(B)
\end{eqnarray}
where the vector ${\bf y}_1$ must verify
\begin{equation}
\label{y1}
{\bf y}_1 \cdot \delta {\bf r}=\dfrac{d}{dt}\left( \sum\limits_{\nu=1}^\kappa \mu_\nu {\bf C}_\nu \cdot \delta {\bf r} \right).
\end{equation}
For $C={\Bbb O}$ (null square matrix of order $N$) the cases $(A)$ and $(B)$ are the same with $A=B$ and the corresponding equations (\ref{eq1}), (\ref{eq2}) do coincide.
As in the previous case, it is necessary to link the elements of $B$, $C$ to the functions $\Psi_\nu$ and their derivatives. Regarding (\ref{y1}), we could say that the condition does not precisely define ${\bf y}_1$ therefore the equations are not closed; actually, the not explicit form of ${\bf y}_1$ in 
(\ref{eq2}) except through $\delta {\bf r}$ makes the equations unusable unless other considerations are added. However, we are interested in the case when both conditions (\ref{delta1}) and (\ref{delta2}) hold and this allows us to eliminate ${\bf y}_1$, as we will see: basically, the physically interesting situation that we will consider corresponds to $A=C$, which allows us to suppress ${\bf y}_1$. 

\noindent
We finally estabilish a relation between the two expressions $A \delta {\bf r}$ and $B \delta {\bf r}+C \delta {\dot {\bf r}}$ (apppearing in (\ref{delta1}) and (\ref{delta2})) written in a way so that the difference appearing in  (\ref{w}):

\begin{equation}
	\label{transpgen}
	B \delta {\bf r}+C \delta {\dot {\bf r}}-\dfrac{d}{dt}(A \delta {\bf r})
	=A \left(\delta {\dot {\bf r}}-\dfrac{d}{dt}\delta {\bf r}\right)
	+\left(B-\dfrac{dA}{dt}\right)\delta {\bf r}+(C-A)\delta {\dot {\bf r}}
\end{equation}
The check of (\ref{transpgen}) is immediate and we refer to it as the {\it transpositional relation}. In examining and employing  (\ref{transpgen}) one does not necessarily have to assume that the two conditions (\ref{delta1}) and (\ref{delta2}) hold simultaneously (that is both the expressions vanish); rather, 
the right terms in the equality (\ref{transpgen}) allow us to investigate about the compatibility and the properties of the two classes of displacements (\ref{delta1}) and (\ref{delta2}).

\subsection{The case $A=C$}

\noindent
Let us consider the special case
\begin{equation}
	\label{a=c}
	A=C
\end{equation}
which will correspond to a significant situation, as we will see.

\begin{prop}
Assume that the displacements verify both (\ref{delta1}) and (\ref{delta2}). If 
(\ref{a=c}) holds, then the vector in (\ref{y1}) can be taken as
\begin{equation}
\label{y=0}
{\bf y}_1={\bf 0}
\end{equation}
and equations (\ref{eq1}) and (\ref{eq2}) are equivalent.
\end{prop}

\noindent
{\bf Proof}. The null vector (\ref{y=0}) satisfies the requirement (\ref{y1}), since (\ref{delta1}) holds with $C=A$.
Moreover, if (\ref{delta1}) and (\ref{delta2}) are both in effective, relation \ref{transpgen}) reduces to 
\begin{equation}
	\label{transpgen1}
	0= {\bf A}_\nu \cdot \left(\delta {\dot {\bf r}}-\dfrac{d}{dt}\delta {\bf r}\right)
	+\left({\bf B}_\nu -\dfrac{d}{dt}{\bf A}_\nu \right)\cdot \delta {\bf r} \qquad \nu=1, \dots, \kappa
\end{equation}
or equivalently, by (\ref{w}) and (\ref{a=c}),
$$
{\bf C}_\nu \cdot W \delta {\bf r}
+\left({\bf B}_\nu -\dfrac{d}{dt}{\bf C}_\nu \right)\cdot \delta {\bf r}=
\left({\bf B}_\nu - \dfrac{d}{dt}{\bf C}_\nu+ W^T {\bf C}_\nu \right)\cdot \delta {\bf r}=0 \qquad \nu=1, \dots, \kappa.
$$
Since the vectors ${\bf A}_\nu$, $\nu=1, \dots, \kappa$, are linearly independent and the set of displacements $\delta {\bf r}$ coincides with the orthogonal complement of the space generated by same vectors, one can write
$$
{\bf B}_\nu - \dfrac{d}{dt}{\bf C}_\nu+ W^T {\bf C}_\nu =\sum\limits_{\sigma=1}^\kappa \rho_\sigma {\bf A}_\sigma
$$
for some $\kappa$--uple $(\rho_1, \dots, \rho_\kappa)$. We conclude that, once again by virtue of (\ref{a=c}), equations (\ref{eq2}) (second version) can be written as 
$$
{\dot {\bf Q}}-{\bf F}=
\sum\limits_{\nu, \sigma =1}^\kappa \left(\mu_\nu \rho_\sigma - {\dot \mu}_\sigma \right){\bf A}_\sigma
$$
which are equivalent to (\ref{eq1}), simply with a different role of the multipliers.
$\quad \square$

\begin{prop}
Again in the case (\ref{a=c}) assume that (\ref{delta1}) and (\ref{delta2}) are valid simultaneously. Then
\begin{equation}
\label{zero1}
{\bf C}_\nu \cdot \left(\delta {\dot {\bf r}}-\dfrac{d}{dt}\delta {\bf r}\right)={\bf 0}
\end{equation}
if and only if
\begin{equation}
\label{zero2}
	{\bf B}_\nu - \dfrac{d}{dt}{\bf A}_\nu = \sum\limits_{\sigma=1}^\kappa \varrho_\sigma {\bf A}_\sigma, \qquad \nu=1, \dots, \kappa
\end{equation}
for some real coefficients $\varrho_1$, $\dots$, $\varrho_\kappa$.
\end{prop}

\noindent
{\bf Proof}.
If (\ref{zero1}) is true, also $({\bf B}_\nu -\frac{d}{dt}{\bf A}_\nu)\cdot \delta {\bf r}=0$ hence (\ref{zero2}) must be verified for some coefficients $\varrho_1$, $\dots$, $\varrho_\kappa$. Conversely, if (\ref{zero2}) holds, then 
$$
\left({\bf B}_\nu -\dfrac{d}{dt}{\bf A}_\nu \right)\cdot \delta {\bf r}=
\sum\limits_{\sigma=1}^\kappa \varrho_\sigma {\bf A}_\sigma \cdot \delta {\bf r}=0
$$
by virtue of (\ref{delta1}) and (\ref{a=c}). We conclude that (\ref{zero1}) is true, due to (\ref{transpgen1}).  $\quad \square$

\begin{cor}
A necessary condition for the commutation rule (\ref{dtdelta}) is the equality (\ref{zero2}).
\end{cor}
Indeed, in that case (\ref{zero1}) holds, which implies (\ref{zero2}). Notice that the inverse statement is not true, that is if (\ref{zero2}) is valid, the rule (\ref{dtdelta}) does not necessarily have to hold.

\subsection{Lagrangian coordinates}

\noindent
In terms of generalized coordinates ${\bf q}=(q_1, \dots, q_n)$ and generalized velocities ${\dot {\bf q}}=({\dot q}_1, \dots, q_n)$, $n=3N$ one has
\begin{equation}
	\label{rq}
	{\bf r}={\bf r}({\bf q},t), \qquad {\dot {\bf r}}({\bf q}, {\dot {\bf q}},t)=
	\sum\limits_{j=1}^n \dfrac{\partial {\bf r}}{\partial q_j}{\dot q}_j+\dfrac{\partial {\bf r}}{\partial q_j}
\end{equation}
where the jacobian matrix 
$\dfrac{\partial {\bf r}}{\partial {\bf q}}=\left( \begin{array}{ccc}
\frac{\partial {\bf r}_1}{\partial q_1} & \dots & \frac{\partial {\bf r}_1}{\partial q_n} \\
\dots & \dots & \dots \\
\frac{\partial {\bf r}_N}{\partial q_1} & \dots & \frac{\partial {\bf r}_N}{\partial q_n}
\end{array} \right) 
$
has maximum rank $n$.

\begin{rem}
When passing from the coordinates ${\bf r}$ to the generalized coordinates ${\bf q}$ we can assume either that  
no further geometric constraint occurs, i.~e.~$n=3N$, or that additional holonomic conditions are juxtaposed: in the latter case, the selection of a smaller number of generalized coordinates $n<3N$ does not lead to substantial changes in the reformulation of the problem (\ref{dap}).
\end{rem}

\noindent
The displacements are expressed in terms of lagrangian variables as

\begin{eqnarray}
		\label{deltar}
			\delta {\bf r}&=&\sum\limits_{j=1}^{n}\dfrac{\partial {\bf r}}{\partial q_j}\delta q_j\\
	\label{deltadotr}
	\delta {\dot {\bf r}}&=&
	\sum\limits_{j=1}^n \dfrac{\partial {\dot {\bf r}}}{\partial q_j}\delta q_j +
	\sum\limits_{j=1}^n \dfrac{\partial {\dot {\bf r}}}{\partial {\dot q}_j}\delta {\dot q}_j=
	\sum\limits_{j=1}^n \dfrac{\partial {\dot {\bf r}}}{\partial q_j}\delta q_j +
	\sum\limits_{j=1}^n \dfrac{\partial {\bf r}}{\partial q_j}\delta {\dot q}_j
\end{eqnarray}
and the problem (\ref{dap}) joined with (\ref{delta1}) or with (\ref{delta2}) is reformulated in the following way:
\begin{equation}
	\label{aorb}
\begin{array}{ll}
(A)\;\;\left\{
	\begin{array}{l}
\sum\limits_{j=1}^n({\dot Q}^{(j)}- F^{(j)})\delta q_j=0 \\
\\
\sum\limits_{j=1}^n \alpha_{\nu,j}\delta q_j=0,\quad  \nu=1, \dots, \kappa
\end{array}
\right.
&
or\;(B)\;\;
\left\{
\begin{array}{l}
	\sum\limits_{j=1}^n({\dot Q}^{(j)}- F^{(j)})\delta q_j=0 \\
	\\
	\sum\limits_{j=1}^n (\beta_{\nu,j}\delta q_j+\gamma_{\nu,j}\delta {\dot q}_j)=0,\quad  \nu=1, \dots, \kappa
\end{array}
\right.
\end{array}
\end{equation}
where
\begin{itemize}
\item[$\bullet$] $v^{(j)}={\bf v}\cdot \dfrac{\partial {\bf r}}{\partial q_j}$, $j=1, \dots, n$, indicates
the $j$--lagrangian component of a $3N$--vector ${\bf v}$; in particular, the relation
${\dot {\bf Q}}^{(j)}=\dfrac{d}{dt}\dfrac{\partial T}{\partial {\dot q}_j}-\dfrac{\partial T}{\partial q_j}$ is well konwn for each $j=1, \dots, n$ and that the presence of active forces coming from a generalized potential, that is
$$
F^{(j)}_P=\dfrac{\partial U}{\partial q_j}-\dfrac{d}{dt}\dfrac{\partial U}{\partial {\dot q_j}}, \qquad 
j=1, \dots, n
$$
for some function $U({\bf q}, {\dot {\bf q}},t)$ makes us write the upper set of equations in $(A)$ or in $(B)$ as
\begin{equation}
	\label{dapl}
	\sum\limits_{j=1}^n \left(
	\dfrac{d}{dt}\dfrac{\partial {\cal L}}{\partial {\dot q}_j}-\dfrac{\partial {\cal L}}{\partial q_j}-F^{(j)}_{NP}\right) \delta q_j=0
\end{equation}
where ${\cal L}({\bf q}, {\dot {\bf q}}, t)=T+U$ is the Lagrangian function and the term $F^{(j)}_{NP}$ takes into account the remaining active forces not deriving from a potential;

\item[$\bullet$]
the coefficients are related to those appearing in (\ref{delta1}) and (\ref{delta2}) by means of 
\begin{equation}
	\label{alphabetagamma}
	\begin{array}{lll}
\alpha_{\nu,j}={\bf A}_\nu \cdot \dfrac{\partial {\bf r}}{\partial q_j},& 
\beta_{\nu,j} ={\bf B}_\nu \cdot \dfrac{\partial {\bf r}}{\partial q_j}+{\bf C}_\nu \cdot \dfrac{\partial {\dot {\bf r}}}{\partial q_j}, & 
		\gamma_{\nu,j}={\bf C}_\nu \cdot \dfrac{\partial {\bf r}}{\partial q_j}
\end{array}
\end{equation}
and they depend on the variables $({\bf q}$, ${\dot {\bf q}},t)$ by virtue of the replacements (\ref{rq}).
\end{itemize}

\noindent
In (\ref{dapl}) we recognize the ordinary way to formulate the D'Alembert principle in generalized coordinates version. 

\noindent
Equations (\ref{eq1}) (case $(A)$) converted to lagrangian variables are immediate:
\begin{equation}
	\label{eq1lagr}
\dfrac{d}{dt}\dfrac{\partial {\cal L}}{\partial {\dot q}_j}-\dfrac{\partial {\cal L}}{\partial q_j}-F^{(j)}_{NP}
	= \sum\limits_{\nu=1}^\kappa \lambda_\nu \alpha_{\nu,j} \quad j=1, \dots, n 
\qquad case\; (A)
\end{equation}
Equations (\ref{eq1lagr}) can be achieved $(i)$ either using the same technique of linear spaces as performed in the previous analysis, $(ii)$ or by calculating the Lagrangian components (by means of the scalar product $\cdot \frac{\partial {\bf r}}{\partial q_j}$, $j=1, \dots, n$) of (\ref{eq1}) and (\ref{eq2}). 
In regard to $(i)$, we see that the matrix $(\alpha_{\nu,j})$ has maximum rank $\kappa$, since $(\alpha_{\nu,j})=A(J_{\bf r}{\bf q})$ (see (\ref{delta1})), hence the formal procedure is identical: the equations $(A)$ in (\ref{aorb}) imply that it must be ${\dot Q}^{(j)}- F^{(j)}=\sum\limits_{\nu=1}^\kappa \lambda_\nu \alpha_{\nu,j}$ for suitable coefficients, hence (\ref{eq1lagr}).

\noindent
Concerning case $(B)$, we start by stating the following
\begin{propr}
The commutation property $\delta {\dot {\bf r}}=\dfrac{d}{dt} \delta {\bf r}$ holds if and only if 
\begin{equation}
	\label{commlagr}
	\dfrac{d}{dt} \left(\delta q_j \right)=\delta \left(\dfrac{d}{dt}q_j \right)
\end{equation} 
holds for any set of independent lagrangian coordinates ${\bf q}=(q_!, \dots, q_n)$.
\end{propr}

\noindent
{\bf Proof}: We see that 
\begin{equation}
	\label{deltaqj}
	\delta {\dot {\bf r}}-\dfrac{d}{dt} \delta {\bf r}=
	\sum\limits_{j=1}^n \left(
	\cancel{\dfrac{\partial {\dot {\bf r}}}{\partial q_j}\delta q_j}
	+\dfrac{\partial {\bf r}}{\partial q_j}\delta {\dot q}_j\right)-
	\sum\limits_{j=1}^n \left(
	\dfrac{\partial {\bf r}}{\partial q_j}\dfrac{d}{dt}(\delta q_j)+
	\cancel{\dfrac{\partial {\dot {\bf r}}}{\partial q_j}\delta q_j}\right)=
	\sum\limits_{j=1}^n\dfrac{\partial {\bf r}}{\partial q_j}\left(\delta {\dot q}_j-\dfrac{d}{dt}\delta q_j\right)
\end{equation}
Since the vectors $\frac{\partial {\bf r}}{\partial q_j}$, $j=1, \dots, n$ are independent, the commutation (\ref{dtdelta}) holds if and only if (\ref{commlagr}) holds. $\quad \square$

\noindent
Assumption (\ref{w}) can be assigned in terms of lagrangian coordinates as
\begin{equation}
	\label{omega}
	\delta {\dot q}_j-\dfrac{d}{dt} \left(\dfrac{d}{dt}q_j \right)=\sum\limits_{h=1}^n \omega_{j,h}\delta q_h.
\end{equation}
Owing to (\ref{deltaqj}), the terms $\omega_{j,h}({\bf q}, {\dot {\bf q}},\bfq2dot^{..},t)$, $j,h=1, \dots, n$, are related to the entries $w_{r,s}$, $r,s=1, \dots, 3N$ of $W$ in (\ref{w}) by means of 
$$
\sum\limits_{s=1}^{3N} w_{r,s}\dfrac{\partial \xi_s}{\partial q_j}=
\sum\limits_{h=1}^n\dfrac{\partial \xi_r}{\partial q_h}\omega_{h,j}\quad r=1, \dots, 3n,\;\; j=1, \dots, n, \quad (\xi_1, \dots, \xi_{3N})={\bf r}
$$
that is, in terms of $W$ and the  matrix $\Omega=(\omega_{j,h})_{j,h=1, \dots, n}$:
\begin{equation}
	\label{womega}
W\dfrac{\partial {\bf r}}{\partial {\bf q}}=\dfrac{\partial {\bf r}}{\partial {\bf q}}\Omega, \qquad \left(\dfrac{\partial {\bf r}}{\partial {\bf q}}\right)_{(r,j)}=\dfrac{\partial \xi_r}{\partial q_j}, \;\;\;r=1, \dots, 3N, \;\;j=1, \dots, n.
\end{equation}
Equations (\ref{eq2}) in terms of the coefficients (\ref{alphabetagamma}) are
\begin{eqnarray}
	\label{eq2lagr}
\dfrac{d}{dt}\dfrac{\partial {\cal L}}{\partial {\dot q}_j}-\dfrac{\partial {\cal L}}{\partial q_j}-F^{(j)}_{NP}
	&=& \sum\limits_{\nu=1}^\kappa \left( \mu_\nu \beta_{\nu,j}-
	\dfrac{d}{dt} \left(\mu_\nu \gamma_{\nu,j} \right)+\mu_\nu\sum\limits_{h=1}^n\gamma_{\nu,h}\omega_{h,j}\right)+y_j
	\\
	\nonumber 
	&=& \sum\limits_{\nu=1}^\kappa \mu_\nu \left( \beta_{\nu,j}-\dfrac{d}{dt} (\gamma_{\nu,j})
	+\sum\limits_{h=1}^n\gamma_{\nu,h}\omega_{h,j} \right)-\sum\limits_{\nu=1}^\kappa {\dot \mu}_\nu \gamma_{\nu,j}+y_j
	\qquad j=1, \dots, n 
\qquad case\;(B)
\end{eqnarray}
The terms containing $\omega_{h,j}$ come from (see (\ref{womega}))
$$
W^T{\bf C}_\nu\cdot \dfrac{\partial {\bf r}}{\partial q_j}
={\bf C}_\nu \cdot W\dfrac{\partial {\bf r}}{\partial q_j}
={\bf C}_\nu \cdot \dfrac{\partial {\bf r}}{\partial {\bf q}}\left(\begin{array}{c} \omega_{1,j}\\
\dots\\
\omega_{n,j} \end{array} \right) 
=\left(\dfrac{\partial {\bf r}}{\partial {\bf q}}\right)^T{\bf C}_\nu \cdot \left(\begin{array}{c} \omega_{1,j}\\
	\dots\\
	\omega_{n,j} \end{array} \right) =
 \left(\begin{array}{c} \gamma_{\nu,1}\\
	\dots\\
	\gamma_{\nu,n} \end{array} \right) \cdot \left(\begin{array}{c} \omega_{1,j}\\
	\dots\\
	\omega_{n,j} \end{array} \right)
$$
and $y_j={\bf y}_1\cdot \dfrac{\partial {\bf r}}{\partial q_j}$ (see (\ref{y1})) is such that
\begin{equation}
\label{yj}
y_j\delta q_j = \dfrac{d}{dt}\left(\sum\limits_{\nu=1}^\kappa \gamma_{\nu,j} \delta q_j\right), 
\qquad j=1, \dots, n.
\end{equation}
It should be noticed that despite the non--complete formal adherence between the Greek and Latin functions in (\ref{alphabetagamma}) (in fact in $\beta_{\nu,j}$ there is an addtional ${\bf C}_\nu$), the formal correspondence between the newtonian eqations (\ref{eq2}) and the lagrangian ones (\ref{eq2lagr}) is respected, in the sense that ``greek'' and ``latin'' terms are present in the same role. This is true by virtue of the cancellation which occurs calculating the following lagrangian components in (\ref{eq2}):
$$
{\bf B}_\nu \cdot \dfrac{\partial {\bf r}}{\partial q_j}- \dfrac{d}{dt}({\bf C}_\nu) \cdot \dfrac{\partial {\bf r}}{\partial q_j}
=\left(\beta_{\nu,j}- \cancel{{\bf C}_\nu \cdot \dfrac{\partial {\dot {\bf r}}}{\partial q_j}}\right)-
\left(\dfrac{d}{dt}\left( {\bf C}_\nu \cdot \dfrac{\partial {\bf r}}{\partial q_j}\right) - \cancel{{\bf C}_\nu \cdot \dfrac{\partial {\dot {\bf r}}}{\partial q_j}}\right)=\beta_{\nu,j}-\dfrac{d}{dt}\gamma_{\nu,j}.
$$
The same formal analogy is present also if we write the transpositional relation (\ref{transpgen}) using lagrangian coordinates, which is 
\begin{equation}
	\begin{array}{l}
		\label{transplagr}
		\sum\limits_{j=1}^n \left( \beta_{\nu,j}\delta q_j+\gamma_{\nu,j}\delta {\dot q}_j\right)-
		\dfrac{d}{dt}\left(\sum\limits_{j=1}^n\alpha_{\nu,j}\delta q_j\right)\\
		\\
		=\sum\limits_{j=1}^n \alpha_{\nu,j}\left(\delta {\dot q}_j-\dfrac{d}{dt}\delta q_j\right)
		+\sum\limits_{j=1}^n \left(\beta_{\nu,j}-\dfrac{d \alpha_{\nu,j}}{dt}\right)\delta q_j
		+\sum\limits_{j=1}^n \left(\gamma_{\nu,j}-\alpha_{\nu,j}\right)\delta {\dot q}_j
	\end{array}
\end{equation}
The relation can be obtained either rearranging directly the terms in the left--hand side, or replacing (\ref{deltar}), (\ref{deltadotr}), (\ref{alphabetagamma}) in (\ref{transpgen}); even in this case, 
cancellations and transfers of terms make the formal structure of (\ref{transplagr}) in agreement with (\ref{transpgen}).

\noindent
Finally, we see that the case $A=C$ (see (\ref{a=c})) opportunely overlaps the following relations:
\begin{equation}
	\label{alpha=gamma}
	\alpha_{\nu,j}=\gamma_{\nu,j}, \qquad \nu=1, \dots, \kappa, \;\;j=1, \dots, n
\end{equation}
and the effects are the same as those outlined in Paragraph . In particular:
\begin{itemize}
	\item[$\bullet$]
	the terms $y_j$ for each $j=1,	\dots, n$ in (\ref{eq2lagr}) can be eliminated,
	\item[$\bullet$]
	the last summation in (\ref{transplagr}) vanishes.
\end{itemize}

\section{A significant application}

\noindent
The correlation between the matrices $A$, $B$, $C$ and the constrints functions (\ref{vinc}) is now specified as follows:

\begin{equation}
	\label{rc}
	{\bf A}_\nu ({\bf r}, {\dot {\bf r}},t)={\bf C}_\nu  ({\bf r}, {\dot {\bf r}},t)=
	\nabla_{\dot {\bf r}}\Psi_\nu, \qquad 
	{\bf B}_\nu  ({\bf r}, {\dot {\bf r}},t)=\nabla_{\bf r}\Psi_\nu
\end{equation}
where $\nabla_{\bf r}=(\frac{\partial}{\partial \xi_1}, \dots, \frac{\partial}{\partial \xi_{3N}})$, setting ${\bf r}=(\xi_1, \dots, \xi_{3N})$. It is evident that the assignment (\ref{rc}) satisfies the requirement (\ref{a=c}).

We indicate by $\delta^{(v)}$ and $\delta^{(c)}$ respectively the operations appearing to (\ref{delta1}) and (\ref{delta2}):
\begin{equation}
	\label{deltavc}
	\delta^{(c)}\Psi_\nu = \nabla_{\dot {\bf r}}\Psi_\nu \cdot \delta {\bf r}, \qquad
	\delta^{(v)}\Psi_\nu = 
	\nabla_{\bf r} \Psi_\nu \cdot \delta {\bf r}+
	\nabla_{\dot {\bf r}}\Psi_\nu\cdot \delta {\dot {\bf r}}
\end{equation}
Definitions (\ref{deltavc}) discover the virtual displacement condition according to the  ${\check {\rm C}}$etaev theory (first equation) and to the vakonomic model (second one).

\noindent
The corresponding equations of motion (\ref{eq1}) and (\ref{eq2}) are now

\begin{equation}
	\label{eq1expl}
	{\dot {\bf Q}}={\bf F}+\sum\limits_{\nu=1}^\kappa \lambda_\nu \nabla_{\dot {\bf r}}\Psi_\nu
\end{equation} 

\begin{eqnarray}
	\label{eq2expl}
	{\dot {\bf Q}}-{\bf F}&=&\sum\limits_{\nu=1}^\kappa \left( \mu_\nu \nabla_{\bf r}\Psi_\nu
	-\dfrac{d}{dt}(\mu_\nu \nabla_{\dot {\bf r}}\Psi_\nu)+\mu_\nu W^T \nabla_{\dot {\bf r}}\Psi_\nu \right)\\
	&=&
	\nonumber
	\sum\limits_{\nu=1}^\kappa \mu_\nu \left( \nabla_{\bf r}\Psi_\nu - \dfrac{d}{dt}(\nabla_{\dot {\bf r}}\Psi_\nu)
	+ W^T \nabla_{\dot {\bf r}}\Psi_\nu\right)
	-\sum\limits_{\nu=1}^\kappa {\dot \mu}_\nu \nabla_{\dot {\bf r}}\Psi_\nu+{\bf y}_1
\end{eqnarray}
where the vector ${\bf y}_1$ is such that (see (\ref{y1}))
$$
{\bf y}_1 \cdot \delta {\bf r}=
\dfrac{d}{dt}\left( \sum\limits_{\nu=1}^\kappa \mu_\nu \nabla_{\dot {\bf r}}\Psi_\nu \cdot \delta {\bf r} \right).
$$
The transpositional relation (\ref{transpgen}) takes the form
\begin{equation}
	\label{transprc}
	\delta^{(v)}\Psi_\nu-\dfrac{d}{dt}\left(\delta^{(c)}\Psi_\nu\right) = \nabla_{\dot {\bf r}}\Psi_\nu
	\cdot \left(\delta {\dot {\bf r}}-\dfrac{d}{dt}\delta {\bf r}\right)-{\cal D}^{(\bf r)}\Psi_\nu \cdot \delta {\bf r}
\end{equation}
where 
\begin{equation}
	\label{derlagr}
	{\cal D}^{(\bf r)}F = \dfrac{d}{dt}
	\left( \nabla_{\dot {\bf r}}F \right) - \nabla_{\bf r}F	
\end{equation}
is the lagrangian derivative of a function $F({\bf r}, {\dot {\bf r}},t)$.

\subsection{Displacements and constrained systems}
	
\noindent
Let us now place the constraints (\ref{vinc}) into the lagrangian formalism: we consider
\begin{equation}
	\label{gvinc}
	g_\nu({\bf q}, {\dot {\bf q}},t)=\Psi_\nu ({\bf r}({\bf q,t}), {\dot {\bf r}}({\bf q}, {\dot {\bf q}},t),t)=0, \qquad \nu=1, \dots, \kappa
\end{equation}
which correspond to (\ref{vinc}) rewritten via (\ref{rq}).
The jacobian matrix  $\dfrac{\partial {\bf g}}{\partial {\dot {\bf q}}}=
\dfrac{\partial {\bf \Psi}}{\partial {\dot {\bf r}}}\dfrac{\partial {\dot {\bf r}}}{\partial {\dot {\bf q}}}=
\dfrac{\partial {\bf \Psi}}{\partial {\dot {\bf r}}}\dfrac{\partial {\bf r}}{\partial {\bf q}}$, where ${\bf g}=(g_1, \dots, g_k)$, has full rank $\kappa$, by virtue of (\ref{psiindep}) and the non--singularity of $\partial {\bf r}/\partial {\bf q}$, since the generalized coordinates ${\bf q}$ are independent.

\noindent
The choice (\ref{rc}) corresponds to 
\begin{equation}
\label{rclagr}
\alpha_{\nu,j} = \gamma_{\nu,j}=\dfrac{\partial g_\nu}{\partial {\dot q}_j}, \quad 
\beta_{\nu,j} = \dfrac{\partial g_\nu}{\partial  q_j}, \qquad \nu=1, \dots, \kappa, \;\;j=1, \dots, n
\end{equation}
and that the displacements (\ref{deltavc}) take the form
\begin{equation}
\label{deltavclagr}
\delta^{(c)}\Psi_\nu = \delta^{(c)} g_\nu =\sum\limits_{j=1}^n \dfrac{\partial g_\nu}{\partial {\dot q}_j}\delta q_j, \quad 
\delta^{(v)}\Psi_\nu = \delta^{(v)}g_\nu  = \sum\limits_{j=1}^n \dfrac{\partial g_\nu}{\partial q_j}\delta q_j+
\sum\limits_{j=1}^n \dfrac{\partial g_\nu}{\partial {\dot q}_j}\delta {\dot q}_j, \qquad \nu=1, \dots, \kappa.
\end{equation}
\noindent
In order to verify (\ref{rclagr}) and (\ref{deltavclagr}) it suffices to take into account (\ref{deltar}), (\ref{deltadotr}) and the relations
\begin{equation}
\label{relgrad}
\left\{
\begin{array}{l}
\dfrac{\partial g_\nu}{\partial q_j}=
\sum\limits_{i=1}^n 
\dfrac{\partial \Psi_\nu}{\partial \xi_i}\dfrac{\partial \xi_i}{\partial q_j}+
\sum\limits_{i=1}^n 
\dfrac{\partial \Psi_\nu}{\partial {\dot \xi}_i}\dfrac{\partial {\dot \xi}_i}{\partial q_j}=
\nabla_{\bf r}\Psi_\nu \cdot \dfrac{\partial {\bf r}}{\partial q_j}+
\nabla_{\dot {\bf r}}\Psi_\nu \cdot \dfrac{\partial {\dot {\bf r}}}{\partial q_j}
\\
\\	
\dfrac{\partial g_\nu}{\partial {\dot q}_j}=
\sum\limits_{i=1}^n 
\dfrac{\partial \Psi_\nu}{\partial {\dot \xi}_i}\dfrac{\partial {\dot \xi}_i}{\partial {\dot q}_j}=
\sum\limits_{i=1}^n 
\dfrac{\partial \Psi_\nu}{\partial {\dot \xi}_i}\dfrac{\partial \xi_i}{\partial q_j}
=\nabla_{\dot {\bf r}}\Psi_\nu \cdot \dfrac{\partial {\bf r}}{\partial q_j}, 
\end{array}
\right.
\end{equation}
where we indexed ${\bf r}=(\xi_1, \dots, \xi_{3N})$.
\begin{rem}
In defiance of the check, the conclusion could only be (\ref{deltavclagr})
if we want the operations $\delta^{(c)}$ and $\delta^{(v)}$ to have a meaning independent of the choice of variables. That is, if we had started with generic Lagrangian coordinates ${\bf q}$, the virtual displacements conditions have necessarily the form (\ref{deltavclagr}).
\end{rem}

\noindent
At this point we can summarize the path through the equations and definitions (\ref{vinc}), (\ref{dap}), (\ref{delta1}), (\ref{delta2}), (\ref{aorb}),  (\ref{eq1lagr}), (\ref{eq2lagr}), (\ref{rc}),  (\ref{eq1expl}), (\ref{eq2expl}),  (\ref{gvinc}), (\ref{rclagr}), (\ref{deltavclagr}) by presenting the following scheme:
\begin{eqnarray}
	\label{dalp}
\sum\limits_{j=1}^n({\dot Q}^{(j)}- F^{(j)})\delta q_j=0, \left\{ \begin{array}{l}g_\nu =0 \\ \nu=1, \dots, \kappa \end{array} \right.& 
	\textrm{D'Alembert--Lagrange principle+constraints}\\
\nonumber
\textrm{ ${\check {\rm C}}$etaev condition} \; \swarrow\;\;(A)&  (B)\;\searrow
\; \textrm{vakonomic condition}
\\
\label{dcdv}
	\delta^{(c)} g_\nu =\sum\limits_{j=1}^n \dfrac{\partial g_\nu}{\partial {\dot q}_j}\delta q_j=0&
	\delta^{(v)}g_\nu  = \sum\limits_{j=1}^n \dfrac{\partial g_\nu}{\partial q_j}\delta q_j+
	\sum\limits_{j=1}^n \dfrac{\partial g_\nu}{\partial {\dot q}_j}\delta {\dot q}_j=0\qquad \\
	\nonumber
	\Downarrow\quad \quad \quad \quad \quad &\Downarrow 
\\
\label{eqs}
{\cal D}^{(q_j)}{\cal L}-F^{(j)}_{NP}
= \sum\limits_{\nu=1}^\kappa \lambda_\nu \dfrac{\partial g_\nu}{\partial {\dot q}_j}\quad \quad \quad
&
\left\{
\begin{array}{l}
{\cal D}^{(q_j)}{\cal L}-F^{(j)}_{NP}=
-\sum\limits_{\nu=1}^\kappa \mu_\nu	{\cal D}^{(q_j)}g_\nu
-\sum\limits_{\nu=1}^\kappa {\dot \mu}_\nu\dfrac{\partial g_\nu}{\partial {\dot q}_j}\\
+\sum\limits_{\nu=1}^\kappa \mu_\nu \sum\limits_{h=1}^n \dfrac{\partial g_\nu}{\partial {\dot q}_h}\omega_{h,j}+y_j
\end{array} 
\right.
\end{eqnarray}
where $y_j$ is defined by (\ref{yj}) and vanishes whenever (\ref{dcdv}) first condition holds, 
\begin{equation}
	\label{derlagrqj}
{\cal D}^{(q_j)}f({\bf q}, {\dot {\bf q}},t)=\dfrac{d}{dt}\dfrac{\partial f}{\partial {\dot q}_j}-
\dfrac{\partial f}{\partial q_j}
\end{equation}
is the lagrangian derivative with respect to the variable $q_j$ of a function $f$. 
In (\ref{dalp}) and (\ref{dcdv}) the index $\nu$ takes each of the values $\nu=1, \dots, \kappa$; the letters $(A)$ and $(B)$ refer to the two categories of displacements (\ref{delta1}) and (\ref{delta2}); the equations in (\ref{eqs}) are $n$ for each type, $j=1, \dots, n$.

\noindent
We now transfer the content of Proposition 1 (which is legitimate because we are in the case (\ref{a=c}), as it is evident from(\ref{alpha=gamma}) and (\ref{rclagr})), to state the main result:

\begin{prop}
Let the dynamics of the constrained system governed by the principle (\ref{dalp}), where the displacements 
are related to the constraint functions by (\ref{dcdv}), first equality, or second equality.
Then, the equations of motion are those written in (\ref{eqs}).
Furthermore, assume that the displacements verify both (\ref{dcdv}). Then $y_j$ defined in (\ref{yj}) can be taken as zero and the two groups of equations in (\ref{eqs}) are equivalent.
\end{prop}

\noindent
We link the equations of motion with the transpositional relation we developed through the formulas (\ref{transpgen}), (\ref{transprc}) and (\ref{transplagr}), which we write again, in light of (\ref{rclagr}) and (\ref{deltavclagr}), as
\begin{equation}
	\label{transpfinale}
	\delta^{(v)} g_\nu -\dfrac{d}{dt}\left(\delta^{(c)}g_\nu\right)=
	\sum\limits_{j=1}^n 
	\dfrac{\partial g_\nu}{\partial {\dot q}_j}\left(\delta {\dot q}_j - \dfrac{d}{dt}\delta q_j\right)
	-\sum\limits_{j=1}^n D^{(q_j)} g_\nu \delta q_j
\end{equation}
where the derivative $D^{(q_j)}$ is defined in (\ref{derlagrqj}).
If (\ref{omega}) is assumed to hold, then the relation takes the form
\begin{equation}
	\label{transpfinale2}
	\delta^{(v)} g_\nu -\dfrac{d}{dt}\left(\delta^{(c)}g_\nu\right)=
	\sum\limits_{h,j=1}^n 
	\dfrac{\partial g_\nu}{\partial {\dot q}_j}\omega_{j,h}\delta q_h-\sum\limits_{j=1}^n D^{(q_j)} g_\nu \delta q_j.
\end{equation}

\noindent
We also transfer the content of Proposition 2 to the relation (\ref{transpfinale}) and we state the following

\begin{prop}
Assume that (\ref{dcdv}) are valid simultaneously. Then
	\begin{equation}
		\label{zero1lagr}
		\sum\limits_{j=1}^n \dfrac{\partial g_\nu}{\partial {\dot q}_j}\left( \delta {\dot q}_j-\dfrac{d}{dt}\delta q_j\right)= 0
	\end{equation}
	if and only if
	\begin{equation}
		\label{zero2lagr}
		{\cal D}^{(q_j)}g_\nu = \sum\limits_{\sigma=1}^\kappa \varrho_\sigma \dfrac{\partial g_\sigma}{\partial {\dot q}_j}, \qquad \nu=1, \dots, \kappa
	\end{equation}
for some real coefficients $\varrho_1$, $\dots$, $\varrho_\kappa$.
\end{prop}

\noindent
The proof is identical to that exhibited in Propostion 2. As we remarked through (\ref{commlagr}), the commutation property (.~e.~(\ref{zero1}) vanishes) is independent of the set of variables are chosen. The necessary condition of Corollary 1 for the commutation (\ref{commlagr}) becomes now (\ref{zero2lagr}).
In \cite{taltransp} it is shown that exact constraints (that is $g_\nu= {\dot f}_\nu({\bf q},t)$ for some $f_\nu$) and constraint functions admitting an integrating factor (that is $\phi_\nu ({\bf q},t)g_\nu ={\dot f}_\nu({\bf q},t)$ for some factor $\phi_\nu$) so that the commutation (\ref{commlagr}) is consistent with these categories. On the other hand, even simple examples of linear constraints (i.~e.~$g_\nu=\sum\limits_{j=1}^n \ell_{\nu,j}({\bf q,t}){\dot q}_j+p_\nu({\bf q},t)$) for which (\ref{zero2lagr}) does not hold, so that the commutation cannot be assumed as valid.

\section{Discussion and next investigation}

\noindent
The two conditions $\delta^{(v)} g_\nu=0$ and $\delta^{(c)}g_\nu=0$ express two different points of view and their application involves various points of discussion: wondering whether they are valid separately, both are valid, one or the other should be applied only to specific classes of mechanical systems are all current topics on the subject.

\noindent
The debate about the simultaneous validity of both conditions in (\ref{dcdv}) -- or better their compatibility -- can profitably look at (\ref{transpfinale}), where it is evident that the question intersects the also relevant and controversial issue about the commutative property of the two operators $\delta$ and $d/dt$
(an useful and extensive discussion on this issue can be found in \cite{flan}).

\noindent
First of all a comparison with the motion equations of the same kind present in the literature is necessary. 
Concerning the class $(A)$, equations (\ref{eqs}), first group, do represent the classic equations of motion for nonholonomic systems deriving from the  ${\check {\rm C}}$etaev condition (\ref{dcdv}), first equality. 
This last condition extends in a simple way a fundamental point in the theory of nonholonomic constraints, which consists of assigning to the linear kinematic constraint $g_\nu=\sum\limits_{j=1}^n \ell_{\nu,j}({\bf q,t}){\dot q}_j+p_\nu({\bf q},t)=0$ the displacements $\delta q_j$ such that 
$\sum\limits_{j=1}^n \ell_{\nu,j}\delta q_j=0$ (we refer, among others, to \cite{lemosbook}, \cite{neimark}); clearly $\delta^{(c)}g_\nu =0$ reduces to the previous condition, if $g_\nu$ is linear with respect to the generalized velocities.

\noindent
Although a rigorous derivation from a physical principle or a theoretical validation of the  ${\check {\rm C}}$etaev condition are uncertain (an extensive discussion on this can be found in \cite{flan}), the method $(A)$ shows two remarkable advantages:
\begin{itemize}
	\item[$(i)$] it does not introduce the question of the commutation rule, since $\delta {\dot q}_j$, $j=1, \dots, n$, are absent in the definition $\delta^{(c)}g_\nu =0$ of virtual displacements; equations (\ref{eq1}) of case $(A)$ can be formulated without making any pronouncements on (\ref{dtdelta}), that is the validity or not of (\ref{commlagr}).
	\item[$(ii)$] it leads to the same equations even by switching the theoretical approach from the 
	d'Alembert principle (as we performed in an algebraic way) to the Hamilton principle via a variational approach \cite{runde}, \cite{saletan}.
\end{itemize}

\noindent
Let us move now on to case $(B)$ and make the following distinction:
\begin{itemize}
	\item[$(B1)$] the hypothesis (\ref{dtdelta}), namely $\delta {\dot {\bf r}}=\dfrac{d}{dt}\delta {\bf r}$ is assumed to hold, hence $\dfrac{d}{dt} \delta q_j = \delta {\dot q}_j$, $j=1, \dots, n$, for any set of lagrangian coordinates  (see (\ref{commlagr})),
	\item[$(B2)$] (\ref{dtdelta}) does not hold and the non--commutative operations is expressed by the formulation of (\ref{w}), namely  $\delta {\dot {\bf r}}=\dfrac{d}{dt}\delta {\bf r}=W\delta {\bf r}$
	(where the matrix $W$ needs to be specified), or equivalently (\ref{omega}).
\end{itemize}
The hypothesis $(B1)$ reduces equations (\ref{eqs}), second group, to (we are also assuming both (\ref{dcdv}), hence $y_j=0$)
\begin{equation}
\label{eqsb1}
{\cal D}^{(q_j)}{\cal L}-F^{(j)}_{NP}=
-\sum\limits_{\nu=1}^\kappa \mu_\nu	{\cal D}^{(q_j)}g_\nu
-\sum\limits_{\nu=1}^\kappa {\dot \mu}_\nu\dfrac{\partial g_\nu}{\partial {\dot q}_j}
\end{equation}
and they do coincide with the vakonomic equations of motion which are obtained either by the integration (as in  \cite{flanelusive}, \cite{liromp}) 
$$
\int\limits_{t_0}^{t_1}\sum\limits_{j=1}^n\left({\cal D}^{(q_j)}{\cal L}-F_{NP}^{(j)}\right)\delta q_jdt=0
$$
or by the constrained Hamilton principle (\cite{arnold})
\begin{equation}
	\label{hpr}
\delta \int\limits_{t_0}^{t_1}\left({\cal L}-F_{NP}^{(j)}+\sum\limits_{\nu=1}^\kappa \mu_\nu g_\nu\right)\delta q_jdt=0.
\end{equation}

\begin{rem}
The treatment by integration needs the additional conditions of both ends fixed
	$$
	\left.\delta {\bf r}\right\vert_{t=t_0}=\left.\delta {\bf r}\right\vert_{t=t_1}={\bf 0}
	$$	
(or equivalently on $\delta q_j$)that further specifies the class of displacements. Our method based on a common algebraic technique for both cases $(A)$ and $(B)$ does not require such a request on the displacements, that is the displacements are treated on equal terms.
\end{rem}

\noindent
The vakonomic model has been decisively criticized over the past few years either for the aspect of specific testing problems showing inconsistency (as in \cite{lemos} and in \cite{xuli}) and from the theoretical perspective.  Regarding the latter, we glean two conclusions from the recent literature (as \cite{flanelusive}, \cite{lemos}  \cite{lewis}):
\begin{itemize}
	\item[$\bullet$] the vakonomic equations $(B1)$ do not reproduce the correct physical state even in the case of linear or homogeneous kinematic constraints,
	 	\item[$\bullet$] the first system in (\ref{eqs}) and (\ref{eqsb1}) are equivalent if and only if the constraints are holonomic, that is $g_\nu = {\dot f}_\nu$.
\end{itemize}

\noindent
As far as we understand, the result in Proposition 3 would seem to offer a broader perspective (i.~e.~even more than the holonomic case) regarding the simultaneous validity of models $(A)$ and $(B)$ (possibly reduced to $(B1)$), attributable to the role of multipliers.

\noindent
The case $(B2)$ offers an interesting and recent proposal to reconcile the two cases $(A)$ and $(B)$ through the action of the matrix $W$ or $\Omega$ (defined in (\ref{w}) and (\ref{omega})) which regulates the non--commutativity of the operations.
We mainly refer to \cite{llibre} and \cite{pastore} in order to compare our results with the ones obtained via 
a variational approach, summarized by (\ref{hpr}).
The equations of motion achieved at a first stage in \cite{llibre} (and revisited in \cite{pastore} as ``modified vakonomic method'') correspond, in our notation, to 

\begin{equation}
\label{eqsb2}
{\cal D}^{(q_j)}{\cal L}-F^{(j)}_{NP}=
-\sum\limits_{\nu=1}^\kappa \mu_\nu	{\cal D}^{(q_j)}g_\nu
-\sum\limits_{\nu=1}^\kappa {\dot \mu}_\nu\dfrac{\partial g_\nu}{\partial {\dot q}_j}
+\sum\limits_{\nu=1}^\kappa \mu_\nu \sum\limits_{h=1}^n \dfrac{\partial g_\nu}{\partial {\dot q}_h}\omega_{h,j}+
\sum\limits_{h=1}^n \dfrac{\partial {\cal L}}{\partial {\dot q}_h}\omega_{h,j}, \quad j=1, \dots, n
\end{equation}
and actually they do not match precisely with (\ref{eqs}), second group, because of the terms in (\ref{eqsb2})
$\sum\limits_{h=1}^n \dfrac{\partial {\cal L}}{\partial {\dot q}_h}\omega_{h,j}$. 
\begin{rem}
The presence of the additional sum can be understood, if we think about (\ref{hpr}) and we rearrange (\ref{eqsb2}) in the form 
\begin{equation}
\label{eqsb2d1}
{\cal D}^{(q_j)}_1{\cal L}-F^{(j)}_{NP}=
-\sum\limits_{\nu=1}^\kappa \mu_\nu{\cal D}^{(q_j)}_1g_\nu 
-\sum\limits_{\nu=1}^\kappa {\dot \mu}_\nu\dfrac{\partial g_\nu}{\partial {\dot q}_j}
\qquad j=1, \dots, n
\end{equation}
where ${\cal D}_1^{(j)}$ is the operator 
\begin{equation}
\label{dj1}
{\cal D}_1^{(q_j)} = {\cal D}^{(q_j)}-\sum\limits_{h=1}^n \omega_{h,j}\dfrac{\partial }{\partial {\dot q}_h} \quad j=1, \dots, n.
\end{equation}
\end{rem}
The disagreement between (\ref{eqsb2d1}) and (\ref{eqs}) second group, which can be written by 
means of (\ref{dj1}) as
\begin{equation}
	\label{dj}
{\cal D}^{(q_j)}{\cal L}-F^{(j)}_{NP}=
-\sum\limits_{\nu=1}^\kappa \mu_\nu	
{\cal D}^{(q_j)}_1 g_\nu
-\sum\limits_{\nu=1}^\kappa {\dot \mu}_\nu\dfrac{\partial g_\nu}{\partial {\dot q}_j} \qquad j=1, \dots, n
\end{equation}
is then overcome if the hypothesis 
\begin{equation}
	\label{wp}
	\sum\limits_{h=1}^n \dfrac{\partial {\cal L}}{\partial {\dot q}_h}\omega_{h,j}=0 \;\;\textrm{for each}\;\;j=1, \dots, n
\end{equation}
is assumed.
The same condition (\ref{wp}) plays a special role in \cite{pastore}, in the sense we are going to explain.
The transpositional condition (\ref{transpfinale2}) can be written by (\ref{dj1}) and by assuming both (\ref{dcdv}) simply as 
\begin{equation}
	\label{d10}
0=	\sum\limits_{h,j=1}^n 
\omega_{h,j}\dfrac{\partial g_\nu}{\partial {\dot q}_h}\delta q_j-\sum\limits_{j=1}^n D^{(q_j)} g_\nu \delta q_j
=-\sum\limits_{j=1}^n D_1^{(q_j)}g_\nu\delta q_j.
\end{equation}
The just written relation is used to claim $D_1^{(q_j)}g_\nu=0$ for each $j=1, \dots, n$ singularly and to write equations (\ref{eqsb2d1}) as in \cite{pastore}, recalling \cite{llibre}:
\begin{equation}
\label{eqsb2ridotte}
{\cal D}^{(q_j)}_1{\cal L}-F^{(j)}_{NP}=
-\sum\limits_{\nu=1}^\kappa {\dot \mu}_\nu\dfrac{\partial g_\nu}{\partial {\dot q}_j}
\qquad j=1, \dots, n
\end{equation}
At this point it is evident that (\ref{wp}) makes these equations equivalent to the  ${\check {\rm C}}$etaev equations (\ref{eqs}), first group, simply by setting $\lambda_\nu = -{\dot \mu}_\nu$ (actually ${\cal D}_1^{(q_j)}{\cal L}= {\cal D}^{(q_j)}{\cal L}$ if (\ref{wp}) holds). Hence in \cite{pastore} the assumption (\ref{wp}) is claimed as sufficient condition for the equivalence of the  ${\check {\rm C}}$etaev and vakonomic equations.

\noindent
Let us comment this conclusion: as far as we are concerned, the relation (\ref{dj}) does not imply {\it logically} the single conditions $D_1^{(q_j)}g_\nu=0$, $j=1, \dots, n$; hence, we understand that the latter conditions are {\it imposed} in order to determine $W$, and they are consistent with (\ref{dj}).

\noindent
In any case, the same condition (\ref{wp}) turns out to be essential also for our presentation, not really for the question of the equivalence of the two types of equations (that we assert to exist in general, independently of (\ref{wp})), but for the consistency of the vakonomic equations following the two different deductions of algebraic or variational type.

\noindent
Our interest in analyzing deeper the topics introduced will focus mainly on the following points:
\begin{itemize}
	\item[$\bullet$] to investigate condition (\ref{zero2lagr}), in order to define the class of constraints for which the terms containing $\delta {\dot q}_j-\frac{d}{dt}\delta q_j$ can be eliminated from the transpositional relation (\ref{transpfinale}), owing to (\ref{zero1lagr}), independently of the validity of (\ref{commlagr}),
	
	\item[$\bullet$] still concerning (\ref{transpfinale}), to determine exactly the validity of the model $(B1)$, that is the entire set of functions $g_\nu$ verifying $\sum\limits_{j=1}^n D^{(q_j)}g_\nu \delta q_j=0$ (the sum, not singularly), in particular whether it coincides with the class of functions that admitting integrating factor,
	
	\item[$\bullet$] what and how many conditions allow us to determine the matrix $W$ or $\Omega$ -- defined in (\ref{w}) and (\ref{omega}) -- and whether we actually always arrive at a closed system.
\end{itemize}

\noindent
The last question is a delicate point and it is crucial in order to close the problem, by fixing the $3N\times 3N$ entries of $W$, or the $n\times n$ entries of $\Omega$.
One aspect of the question has already emerged regarding the validity (which we have questioned) of 
${\cal D}^{(q_j)}_1 g_\nu =0$ for each $j=1, \dots, n$: this would provide $\kappa \times n$ conditions on the $n^2$ entries of $\Omega$. 
On the contrary, the (only one) condition (\ref{d10}) supplies just $3N-\kappa$ conditions, from our point of view.

\noindent
Finally, we remark that another source of information for the matrix $W$ or $\Omega$ may come from the particular structure of the problem: for istance, if ${\bar q}_j$ is an ignorable lagrangian coordiate in ${\cal L}$, that is $\frac{\partial {\cal L}}{\partial {\bar q}_j}=0$. Let us add the hypotheses $\frac{\partial g_\nu}{\partial 
{\dot {\bar q}}_j}=0$ for any $\nu=1, \dots, \kappa$ (the circumstance is not uncommon, as in the examples studied in \cite{pastore}): then the corresponding equation of motion (\ref{eqs}), first group states that 
$$
\frac{\partial {\cal L}}{\partial {\dot {\bar q}}_j}={\bar p}_j=costant
$$
and transferring this information to the corresponding equation on the right one gets the condition on $\omega_{h,j}$
$$
{\bar p}_j=\sum\limits_{h=1}^n \left( g_\nu - \sum\limits_{h=1}^\kappa 
\frac{\partial g_\nu}{\partial {\dot q}_h}\omega_{h,j}\right)
$$
On the contrary, if (\ref{eqsb2ridotte}) are claimed to hold, then the deduced condition confirms (\ref{wp}) for the index $j$ corresponding to the ignorable variable ${\bar q}_j$.


\end{document}